\documentclass[reprint,showpacs,amsmath,amssymb,aps]{revtex4}
\usepackage{graphicx,color}

\begin{document}

\title{The Minimal Flavor Structure from Decomposition of the Fermion Mass Matrix}
\author{Ying Zhang \footnote{E-mail: hepzhy@mail.xjtu.edu.cn. Corresponding author. }}
\address{Institute of Theoretical Physics, School of Physics, Xi'an Jiaotong University, Xi'an, 710049, China}

\date{\today}

\begin{abstract}
The minimal flavor structures for both quarks and leptons are proposed to address fermion mass hierarchy and flavor mixings by bi-unitary decomposition of the fermion mass matrix. The real matrix ${\bf M}_0^f$ is completely responsive to family mass hierarchy, which is expressed by a close-to-flat matrix structure. The left-handed unitary phase ${\bf F}_L^f$ provides the origin of CP violation in quark and lepton mixings, which can be explained as a quantum effect between Yukawa interaction states and weak gauge states. The minimal flavor structure is realized by just 10 parameters without any redundancy, corresponding to 6 fermion masses, 3 mixing angles and 1 CP violation in the quark/lepton sector. 
This approach provides a general flavor structure independent of the specific quark or lepton flavor data. We verify the validation of the flavor structure by reproducing quark/lepton masses and mixings. Some possible scenarios that yield the flavor structure are also discussed.
\end{abstract}

\maketitle


\section{Introduction}
Although the standard model (SM) has been proven as a successful  description of strong and electroweak interactions, there are some remaining unknown issues underlying fermion flavor structure \cite{Feruglio:2015jfa,Raidal:2008jk,ZZX2020PR}. 
In the quark sector of the SM, all flavor information is encoded into the complex Yukawa couplings ${\bf y}_{ij}^f$ (for $f=u,d$). 
After electroweak symmetry breaking, quarks obtain masses
${\bf M}_{ij}^f=v_0{\bf y}_{ij}^f/\sqrt{2}$ with Higgs vacuum expectation value (VEV) ${v_0}/{\sqrt{2}}$.
Complex Yukawa couplings ${\bf y}_{ij}^{u}$ and ${\bf y}_{ij}^d$ include $2\times(9+9)$ free parameters. 
However, they correspond to only 10 phenomenological observables: 6 quark masses, 3 CKM mixing angles and 1 CP violation (CPV). The redundancy implies that there must be an unknown flavor structure hidden in the SM Yukawa couplings \cite{2019Zupan}. 
Another insight comes from the similar flavor structure in the quark and lepton sector. 
In the minimal extension lepton sector with 3 Dirac neutrinos, the story of quark flavor is restated: charged lepton and neutrino masses are provided from complex Yukawa terms; the total number of observables is also 10, namely, 6 lepton masses, 3 PMNS mixing angles and 1 Dirac CPV; and a nonvanishing CPV also exists in neutrino mixings. This information inspires us to seek the common flavor structures behind quarks and leptons that give rise to fermion mass hierarchy and flavor mixings in a unified manner.

There are two popular approaches to investigate flavor structure. 
The first approach is started by constructing a new physics model based upon some symmetry or dynamical principle, which is referred to as top-down. Many models and theories have addressed the flavor issue through this approach, including examples based on discrete symmetry, GUT, extra dimension theory and string theory \cite{2016PetcovNPB,Girardi2016NPB,ZZXing2014PLB,Dias2018PRD,King2016RPP,Archer2012JHEP,Nima2000PRD,Bhattacharyya2012PRL,Martinez2018PRD}. 
Extra particles including those of these models, such as flavons, sterile neutrinos and extra Higgs, play roles of flavor symmetry breaking and/or origination of the CPV. With the extension of the SM, the particles, especially around the TeV scale, need be arranged subtly to maintain the balance between new physics predictions and the SM observations. Some challenging and complex experiments have been designed and even conducted in the search for possible effects beyond the SM. Currently, however, there is still no decisive signal.
Another approach is the bottom-up approach, which is inspired by the relation between mass and/or mixing angle values. These experimental data can be interpreted as powers of one or more small flavor breaking parameters, such as bi-maximal mixing (BM) and tri-bi-maximal mixing (TBM) in lepton mixings \cite{2012AltarelliFP,2012ZHZhaoPRD}.

In addition to addressing hierarchal masses and flavor mixings with nonvanishing CPV, an expected flavor structure should provide a common mechanism that functions well in both the quark sector and lepton sector. Furthermore, the mechanism does not depend on the special numerical results. The most desired flavor structure should include only 10 parameters in the Lagrangian corresponding to just 10 observables in the quark/lepton sector, which is the so-called the minimal flavor structure. This struture can provide a phenomenological platform to build flavor models and understand flavor nature.
In the paper, we adopt an approach that is different from top-down or bottom-up strategies. 
Flavor structure is directly analyzed from bi-unitary decomposition of fermion mass matrix. The method is rooted in the success of the SM. 
Currently, almost all 10$+$10 independent parameters in the quark and lepton sectors have been measured, except the Dirac CPV in neutrino mixing. The known data provide a good checkpoint for the validation of possible flavor structure behind fermion mixing and/or mass distribution. 

Starting from the SM fermion mass matrix, it can be decomposed into the production of  a real  matrix ${\bf M}_0^f$ and left- and right-handed unitary matrix ${\bf F}_{L,R}^f$. 
We show that these matrixes fill different roles in flavor physics. ${\bf M}_0^f$ is completely responsible for fermion mass. ${\bf F}_{L}^f$ provides a complex phase in flavor mixing transformation, which causes a CPV  in CKM/PMNS. 
The minimal flavor structure can be realized by parameterizing ${\bf F}_{L,R}^\dag$ and ${\bf M}_0^f$ to address mass hierarchies and flavor mixings. 
Inspired by hierarchal eigenvalues that arise from a flat matrix with nondiagonal corrections, we propose that the family mass hierarchy problem can be resolved by a close-to-flat matrix structure in flavor space. 
The similar structure has also been proposed as a quasi-democratic pattern from discrete symmetry with complicated parameterization \cite{Fritzsch2017CPC,Sogami1998PTP,Fukuura1999PRD}. In the close-to-flat matrix, only three real parameters are needed. As a byproduct, family universal Yukawa coupling can arise naturally \cite{Shinohara1998PTP,DSDu1993MPLA,Branco1995PLB,Teshima1997PTP}; it determines the sum of a family mass.
We also study the role of the unitary matrix ${\bf F}_{L,R}^f$ in flavor mixing. Because of left-handed weak $SU(2)_L$ symmetry, the CKM only requires left-handed ${\bf F}_L^f$ and is blind to right-handed ${\bf F}_R^f$.  
The minimal parameterization of ${\bf F}_L^f$ requires two independent phases. 
We present that ${\bf F}^f_L$ can be explained as a quantum phase between Yukawa interaction states and weak interaction states. 
In contrast to complex couplings from Yukawa terms and/or vacuum structure from new physics, the complex superposition coefficients provide a new way to explain the origin of CP violation.
It enables us to understand the role of CPV in many interesting issues of particle physics and cosmology \cite{Branco2012RMP}.
We also generalize the minimal flavor structure to the lepton sector with the normal hierarchy Dirac neutrinos. All formulas and conclusions can be generalized from quarks to leptons directly.  

The remainder of the paper is organized as follows: in the next section, the SM fermion mass matrixes are decomposed into bi-unitary form. 
The roles and properties of ${\bf M}_0^f$ and ${\bf F}_{L,R}^f$ are analyzed. 
In Sec. III, the minimal parameterization of ${\bf M}_0^f$ is proposed and the physical roles of parameters are analyzed. All quark and lepton masses are reproduced to fit experimental data. In Sec. IV, we focus on CPV in flavor mixing. Only two phases are involved in ${\bf F}_{L,R}^f$. The physical means of these phases are discussed as a quantum superposition effect.
CKM and PMNS are also reproduced successfully, which proves the validity of the mechanism. In Sec. V, we discuss a possible new physics model that can generate the minimal flavor structure.
Finally, a short conclusion and discussion are provided.

\section{Bi-unitary decompositon of the fermion mass matrix}
In the SM, the quark Yukawa term is 
	\begin{eqnarray}
		\mathcal{L}_Y=y^d_{ij}\bar{\Psi}_L^i\Phi{\psi^{d,j}_R}+y^u_{ij}\bar{\Psi}_L^i\tilde{\Phi}{\psi^{u,j}_R}+H.c.
	\end{eqnarray}
with generation index $i,j$, left-handed quark doublet $\Psi_L^i$, right-handed up-type quark $\psi^{u,i}_{R}$, down-type quark ${\psi^{d,i}_R}$ and Higgs doublet $\Phi$.
After electroweak symmetry breaking, the Higgs obtains VEV $\langle {\Phi}\rangle=v_0/\sqrt{2}$ and the quark mass matrix becomes 
$${\bf M}_{ij}^f=\frac{v_0}{\sqrt{2}}{\bf y}_{ij}^f,~~f=u,d.$$ 
Making bi-unitary transformations
\begin{eqnarray}
\psi_{L}^f=({\bf U}^f_{L})^\dag \psi^{f,m}_{L},~~~~
\psi_{R}^f=({\bf U}^f_{R})^\dag \psi^{f,m}_{R},
\label{eq.diagonalizetransf}
\end{eqnarray}
to mass eigenstate $\psi^{f,m}_{L,R}$, ${\bf M}^f$ is diagonalized as
\begin{eqnarray}
{\bf M}^f\rightarrow {\bf U}_L^f{{\bf M}^f}({\bf U}_R^f)^\dag={\rm diag}(m^f_1,m^f_2,m^f_3)
\label{eq.diagM}
\end{eqnarray}
In mass basis, the weak charged current interaction becomes
\begin{eqnarray}
\frac{g}{\sqrt{2}}\bar{\psi}^{u,m}_L\gamma^\mu{\bf U}_{CKM}\psi^{d,m}_L W_\mu^+
			+H.c.
\end{eqnarray}
The CKM matrix is determined by bi-unitary transformations as ${\bf U}^u_L{{\bf U}^d_L}^\dag$.
A conventional parameterization of CKM is \cite{ClauPRL1984}
\begin{eqnarray*}
{\bf U}_{CKM}=\left(\begin{array}{ccc}
		1&0&0\\
		0& c_{23} & s_{23}\\
		0& -s_{23} & c_{23}
		\end{array}\right)
		\left(\begin{array}{ccc}
		c_{13}&0 & s_{13}e^{-i\delta_{CP}}\\
		0 &1 &0\\
		-s_{13}e^{i\delta_{CP}} &0 & c_{13}
		\end{array}\right)
		\left(\begin{array}{ccc}
		c_{12} & s_{12} &0 \\
		-s_{12} & c_{12} &0\\
		0&0&1
		\end{array}\right)
\end{eqnarray*}
with $c_{ij}\equiv\cos\theta_{ij},s_{ij}\equiv \sin\theta_{ij}$. (Alternatively, Wolfenstein parameterization can be defined in \cite{WolfensteinPRL1983}.)

The above formula can also be generalized to the lepton sector with three right-handed Dirac neutrinos. The lepton mixing PMNS matrix has form ${\bf U}_{PMNS}={\bf U}_L^\nu{{\bf U}^e_L}^\dag$. 
Because nonvanishing CPV in both CKM and PMNS has been confirmed by experiments, the complex phase requires complex Yukawa couplings in the SM.

Using eq. (\ref{eq.diagM}), a general  decomposition of ${\bf M}^f$ can be expressed in a bi-unitary form  
\begin{eqnarray}
{\bf M}^f=({\bf {F}}_L^f)^\dag{\bf {M}}_0^f{\bf {F}}_R^f
\label{eq.Mdecomposition}
\end{eqnarray} 
with unitary matrix ${\bf F}_{L,R}^f$.
Here ${\bf M}_0^f$ can be defined from diagonalized mass matrix by a real orthogonal transformation ${\bf U}_0^f$
\begin{eqnarray}
{\bf U}_0^f{\bf {M}}_0^f({\bf U}_0^f)^T={\rm diag}(m_1^f,m_2^f,m_3^f).
\label{eq.quarkU0}
\end{eqnarray}
The bi-unitary transformations in Eq.(\ref{eq.diagonalizetransf}) can be constructed by  
	\begin{eqnarray}
	{\bf U}_L^f= {\bf U}_0^f{{\bf {F}}_L^f},~~~{\bf U}_R^f= {\bf U}_0^f{{\bf {F}}_R^f}.
	\label{eq.doublesideU0}
	\end{eqnarray}
 Thus, the real matrix ${\bf {M}}_0^f$ completely encodes fermion mass information.

In the quark sector, flavor mixings are expressed into
	\begin{eqnarray}
		{\bf U}_{CKM}={\bf U}^u_L{{\bf U}^d_L}^\dag
		={\bf U}^u_{0}{\bf {F}}^u_L({\bf {F}}_L^d)^\dag{{\bf U}^d_{0}}^\dag.
		\label{eq.CKM}
	\end{eqnarray}
The production ${\bf {F}}^u_L({\bf {F}}_L^d)^\dag$ provides the origin of complex phases between different flavors that is required by nonvanishing CPV in CKM.
In the lepton sector, the neutrino mixing matrix has the similar form 
	\begin{eqnarray}
		{\bf U}_{PMNS}={\bf U}^e_L{{\bf U}^\nu_L}^\dag
		={\bf U}_{0}^e{\bf {F}}^e_L({\bf {F}}_L^\nu)^\dag{{\bf U}_{0}^\nu}^\dag.
	\label{eq.PMNS}
	\end{eqnarray}
The leptonic CPV is generated from complex ${\bf {F}}^e_L({\bf {F}}_L^\nu)^\dag$.
After the mass matrix decomposition, we have found that ${\bf M}_0^f$ and unitary matrix ${\bf F}_{L,R}^f$ play different roles with respect to flavor structure:
\begin{itemize}
	\item[(1)] real matrix ${\bf M}_0^f$ completely determines fermion mass eigenvalues; 
	\item[(2)] left-handed ${\bf F}_{L}^f$ provides complex phases, which is a necessary condition for CPV in CKM/PMNS. 
\end{itemize}

The approach can be used to Majorana neutrinos with some improvements. If the neutrino mass matrix is diagonalized by a unitary transformation ${\bf U}^\nu$
		$${\bf U}^\nu {\bf M}^\nu ({\bf U}^\nu)^T={\rm diag}(m_1^\nu,m_2^\nu,m_3^\nu).$$
		Majorana neutrino mass matrix can be decomposed by 
			${\bf M}^\nu=({\bf F}^\nu)^\dag {\bf M}_0^\nu {\bf F}^\nu$
		with ${\bf F}^\nu=({\bf U}_0^\nu)^T({\bf U}^\nu)^*$. Real matrix ${\bf M}_0^\nu$ is diagonalized by real rotation ${\bf U}_0^\nu$. And PMNS matrix for Majorana neutrino becomes
			$${\bf U}_{PMNS}={\bf U}_L^e({\bf U}^\nu)^T={\bf U}_0^e{\bf F}_L^e({\bf F}^\nu)^T({\bf U}_0^\nu)^T.$$
In the paper, we will focus on  Dirac neutrinos.
\section{Family mass hierarchy}
Quarks, charged leptons and even normal hierarchy (NH) neutrinos all exhibit an important characteristic: family mass hierarchy, i.e., $h_{13}^f, h^f_{12}\ll 1$ for $f=u,d,l,\nu$ with hierarchy definition $h^f_{ij}=m^f_i/m^f_j$. 
Due to physical masses as eigenvalues of  ${\bf M}_0^f$,  the family mass hierarchy problem must be tracked from ${\bf M}_0^f$. In this section, we focus on the minimal parameterization of real matrix ${\bf M}_0^f$ and its verification.
\subsection{The minimal parameterization of ${\bf M}_0^f$}
\begin{figure}[htbp]
\begin{center}
\includegraphics[height=0.27 \textheight]{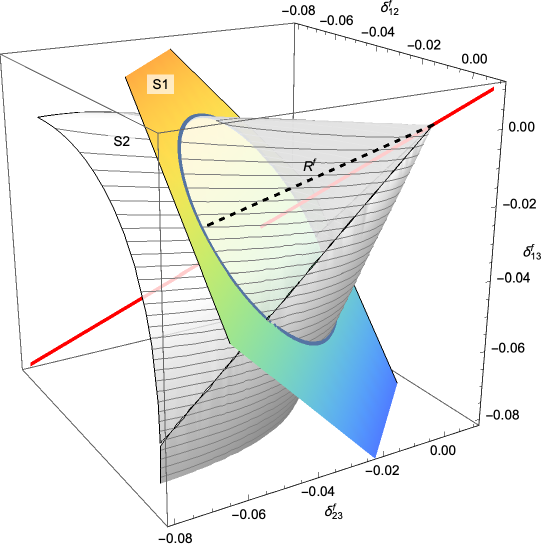}
\caption{{\bf Parameter space and family mass hierarchies.} Surfaces $S1$ (rainbow colors) and $S2$ (gray) are shown at $h_{12}^f=0.05$ and $h_{23}^f=0.02$. The intersection line (deep blue) is allowed by the fermion mass hierarchies. The invariance $R^f$ and axis $(1,1,1)$ are labeled by black dashed and red solid lines, respectively.}
\label{fig.explainVP}
\end{center}
\end{figure}

In mathematics, hierarchical eigenvalues can naturally arise from the close-to-flat matrix. Considering a $2\times2$ matrix with symmetric nondiagonal element
	\begin{eqnarray}
		\left(\begin{array}{cc}1& 1-\delta  \\ 1-\delta & 1 \end{array}\right).
	\end{eqnarray}
$\delta$ stands for a small breaking effect.
Its two eigenvalues $\chi_i$ are
	\begin{eqnarray}
		\chi_1&=&\delta
		\\
		\chi_2&=&2-\delta
	\end{eqnarray}
A very large order $h_{12}$ can be generated by perturbation $\delta$ 
	\begin{eqnarray}
		h_{12}=\frac{\chi_1}{\chi_2}\simeq\frac{\delta}{2}+\mathcal{O}(\delta^2).
	\end{eqnarray}
Inspired by the property, we assume that the fermion mass structure is driven by the same mechanism.
Defining a $3\times3$ matrix
	\begin{eqnarray}
		{\bf I}_\Delta^f=\left(\begin{array}{ccc}1 & 1+\delta_{12}^f & 1+\delta_{13}^f\\ 1+\delta_{12}^f & 1 & 1+\delta_{23}^f \\ 1+\delta_{13}^f & 1+\delta_{23}^f & 1\end{array}\right)
	\end{eqnarray}
with flavor-dependent nondiagonal perturbations $\delta_{ij}^f$,
the real matrix ${\bf M}_0^f$ has a form
	\begin{eqnarray}
		{\bf M}_0^f=\frac{m_{\Sigma}^f}{3}{\bf I}_\Delta^f
	\label{eq.M0}
	\end{eqnarray}
Here, $m_{\Sigma}^f\equiv\sum_i m_i^f$ is total family mass. Due to  flavor symmetry breaking being taken by perturbations $\delta_{ij}^f$, $m_{\Sigma}^f$ hints that a family universal coupling can be introduced as $y^f=\frac{\sqrt{2}}{v_0}\frac{m_\Sigma^f}{3}$.
With $\delta_{ij}^f$ vanishing, the matrix ${\bf I}_\Delta^f$ becomes flat, which means that all flavors are symmetric and that no flavor is prioritized. 
The case in which ${\bf I}_\Delta^f$ has eigenvalues $0, 0, 3$ corresponds with the case in which only the 3rd generation is massive and the first two generations are massless, which happens to coincide with results from Weinberg's models of lepton and quark masses \cite{2020Weinberg}. When flavor symmetry is broken by nonvanishing $\delta_{ij}^f$, the two light flavors obtain small masses. 
Here, only two free parameters are needed to control $m_1^f$ and $m_2^f$. At 1-order approximate, the physical masses are
\begin{eqnarray}
		m^f_{1,2}&=&\frac{y^fv_0}{\sqrt{2}}\left(\frac{1}{3}S^f\mp\frac{2}{3}\sqrt{Q^f}\right)+\mathcal{O}(\delta^2)
		\label{eq.m1m2}\\
		m^f_3&=&\frac{y^fv_0}{\sqrt{2}}\left(3-\frac{2}{3}S^f\right)+\mathcal{O}(\delta^2)
		 \label{eq.m3}
	\end{eqnarray}
with parameters $S^f,Q^f$ as
	\begin{eqnarray}
		S^f&\equiv& -\delta^f_{12}-\delta^f_{23}-\delta^f_{13}
\label{eq.S}\\
		Q^f&\equiv&(\delta^f_{12})^2+(\delta^f_{23})^2+(\delta^f_{13})^2-\delta^f_{12}\delta^f_{23}-\delta^f_{23}\delta^f_{13}-\delta^f_{13}\delta^f_{12}
\label{eq.Q}
	\end{eqnarray}
Obviously, the hierarchy $h_{12}^f$ vanishes when $S^f-2\sqrt{Q^f}$ tends to zero, i.e. $\Big(\sqrt{-\delta_{12}^f}+\sqrt{-\delta_{23}^f}+\sqrt{-\delta_{13}^f}\Big)\rightarrow0$. And the hierarchy $h_{23}^f$ vanished when $S^f+2\sqrt{Q^f}$ tends to zero.
Using eqs. (\ref{eq.m1m2}) and (\ref{eq.m3}), $S^f$ and $Q^f$ can be expressed into hierarchies
\begin{eqnarray}
		S^f&=&\frac{9}{2}(h^f_{23}+h^f_{12}h^f_{23}-(h^f_{23})^2)+\mathcal{O}(h^3)
		\label{eq.hierarchy2SQ1}\\
		Q^f&=&\frac{81}{16}(h^f_{23})^2+\mathcal{O}(h^3)
	\label{eq.hierarchy2SQ2}
\end{eqnarray}
The two quantities $S^f$ and $Q^f$ split family masses, and one of $\delta_{ij}^f$ is free. This can be understood from $SO(2)$ symmetry.
Defining $R^f=(\delta_{12}^f)^2+(\delta_{23}^f)^2+(\delta_{13}^f)^2$, it is $SO(2)$ invariant along the direction of $(1,1,1)$ and is fixed by mass hierarchies
\begin{eqnarray}
	R^f=\frac{1}{3}((S^f)^2+2Q^f)=\frac{81}{4}(h_{23}^f)^2\Big(\frac{1}{2}-\frac{h_{12}^f}{3}\Big)+\mathcal{O}(h^4)
	\label{eq.Rf}
\end{eqnarray} 
Fig. \ref{fig.explainVP} explains the relationship between mass hierarchy and perturbations $\delta_{ij}^f$.  
Eq. (\ref{eq.S}) represents a plane $S1$ in the space $(\delta^f_{12},\delta^f_{23},\delta^f_{13})$. Eq. (\ref{eq.Q}) represents a surface $S2$ that is determined by hierarchy $h_{23}^f$. Their intersection line corresponds to the parameter space allowed by fermion mass data. The intersection forms a circle with invariant distant $R^f$ to the origin.
To explain more details, the free  rotation angle on the surface $S1$ is labelled by $\theta^f$ as shown in Fig. \ref{fig.explainVP}, the perturbations $\delta_{ij}^f$ are determined by $\theta^f$ as 
	\begin{eqnarray*}
		\delta_{12}^f&\simeq&\Big(-\frac{3c_\theta}{4}+\frac{9s_\theta}{4\sqrt{3}}-\frac{3}{2}\Big)h_{23}+\mathcal{O}(h^2)
		\\
		\delta_{23}^f&\simeq&\Big(-\frac{3c_\theta}{4}-\frac{9s_\theta}{4\sqrt{3}}-\frac{3}{2}\Big)h_{23}+\mathcal{O}(h^2)
		\\
		\delta_{13}^f&=&\Big(\frac{3c_\theta}{2}-\frac{3}{2}\Big)h_{23}+\mathcal{O}(h^2)
	\end{eqnarray*}
(The case corresponds to mass eigenvalues $(m_1^f,m_2^f,m_3^f)=(0,3h_{23},3-3h_{23})$ at the leading order of hierarchies.)
For fixed mass hierarchies, the orthogonal rotation ${\bf U}_0^f$ in eq. (\ref{eq.quarkU0}) 
provides a  rotation degree of freedom $\theta^f$ to flavor mixing. The CKM mixing matrix in eq. (\ref{eq.CKM}) is determined by two rotation angles $\theta^u,\theta^d$ from ${\bf U}_0^{u,d}$ and two Yukawa phases $\lambda_1^u$ and $\lambda_1^d$ 
$${\bf U}_{CKM}= \Big({\bf U}_0^u(\theta^u) \Big){\rm diag}\left(1,e^{i\lambda^u_1},e^{i\lambda^u_2}\right)\Big({\bf U}_0^d(\theta^d)\Big)^T.$$ 
In the lepton sector, the PMNS mixing matrix has a similar form.

\subsection{Reproduction of fermion mass hierarchies}
Two of three $\delta_{ij}^f$ are bounded by two hierarchies, and one is free. 
In the order of $\mathcal{O}(\delta_{ij}^f)$, $S^f,Q^f$ is fixed by experiment data in eqs. (\ref{eq.hierarchy2SQ1}) and (\ref{eq.hierarchy2SQ2}). 
We choose $\delta_{13}^f$ as a free parameter, and $\delta_{12}^f,\delta_{23}^f$ can be solved from eq. (\ref{eq.S},\ref{eq.Q}) with fixed mass hierarchy $h_{ij}^f$. There are two kinds of methods to find the parameter space of $\delta_{ij}^f$:
\begin{itemize}
	\item[1.] Parameter space scanning: Initializing three $\delta_{ij}^f$ in the range of $(-S^f,0)$ in terms of eq. (\ref{eq.S}), the eigenvalues of mass matrix eq. (\ref{eq.M0}) can be calculated to match fermion mass data. This method is adopted for the quark sector to investigate the $1\sigma$ range in the quark sector. 
	\item[2.] Solution from hierarchies: For fixed $S^f,Q^f$, the perturbations $\delta_{12}^f$ and $\delta_{23}^f$ can be solved from each input $\delta_{13}^f$ as
	\begin{eqnarray*}
		\delta_{12,23}^f&=&-\frac{\delta^f_{13}+S^f}{2}\pm\frac{1}{2\sqrt{3}}\sqrt{-9(\delta^f_{13})^2+4Q^f-6\delta^f_{13}S^f-(S^f)^2}.
	\end{eqnarray*}
When $\delta_{13}^f$ adopts the entire possible range, the complete parameter space is obtained. We adopt this method in the lepton sector due to the high precision of the charged lepton mass data. 
\end{itemize}
\subsubsection*{Quark mass hierarchies}
\begin{figure}[htbp]
\begin{center}
\includegraphics[height=0.33 \textheight]{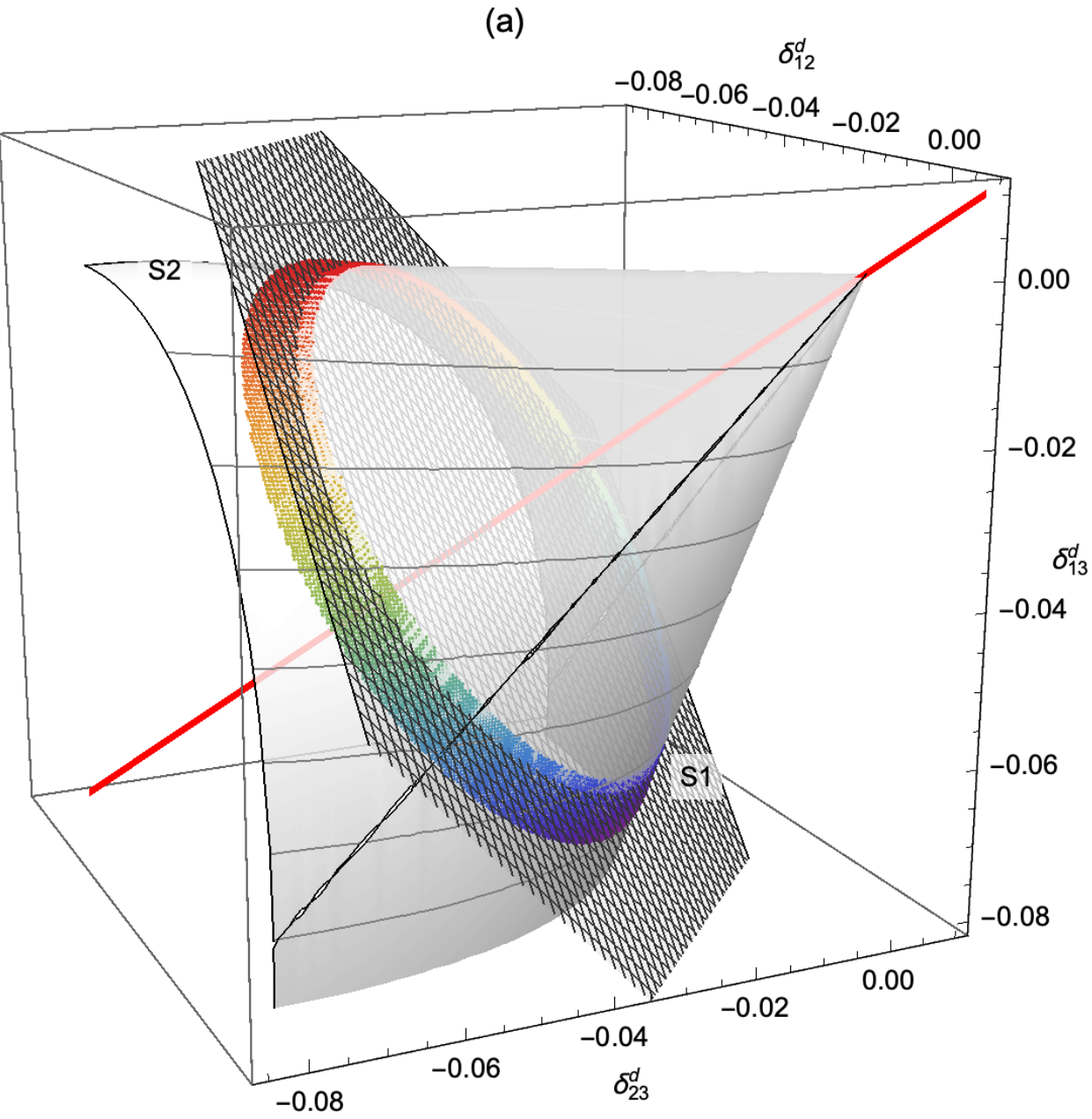}  
\includegraphics[height=0.33 \textheight]{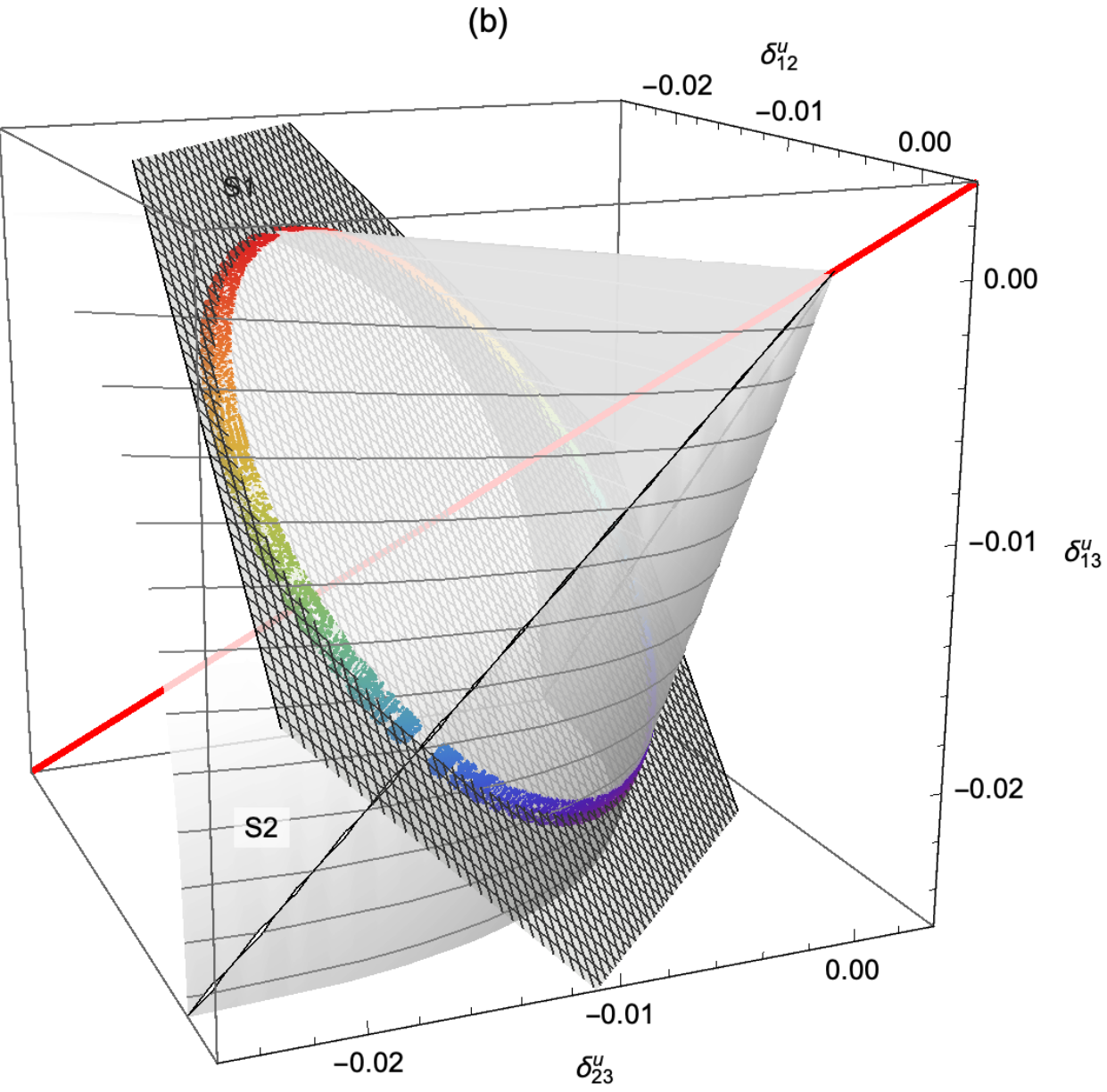}  
\caption{ Parameter spaces $(\delta_{12}^f,\delta_{23}^f,\delta_{13}^f)$ for (a) down-type quarks and (b) up-type quarks in $1\sigma$ CL.}
\label{fig.mdmu}
\end{center}
\end{figure}

The current quark mass data are listed in Tab. IV. 
Setting ${y^fv_0}/{\sqrt{2}}=(m^f_1+m^f_2+m^f_3)/3$, the quark masses can be determined by calculating ${\bf M}_0^f$ eigenvalues by scanning the complete possible range. If mass eigenvalues satisfy the experimental data, the set of perturbations is recorded. 
The numerical results show that all allowed data are distributed on a circle in Fig. \ref{fig.mdmu}. 
The distribution thickness is controlled by the $1\sigma$ range of  $h^u_{23}$. 
\subsubsection*{Lepton mass hierarchies}
\begin{figure}[htbp]
\begin{center}
\includegraphics[height=0.33 \textheight]{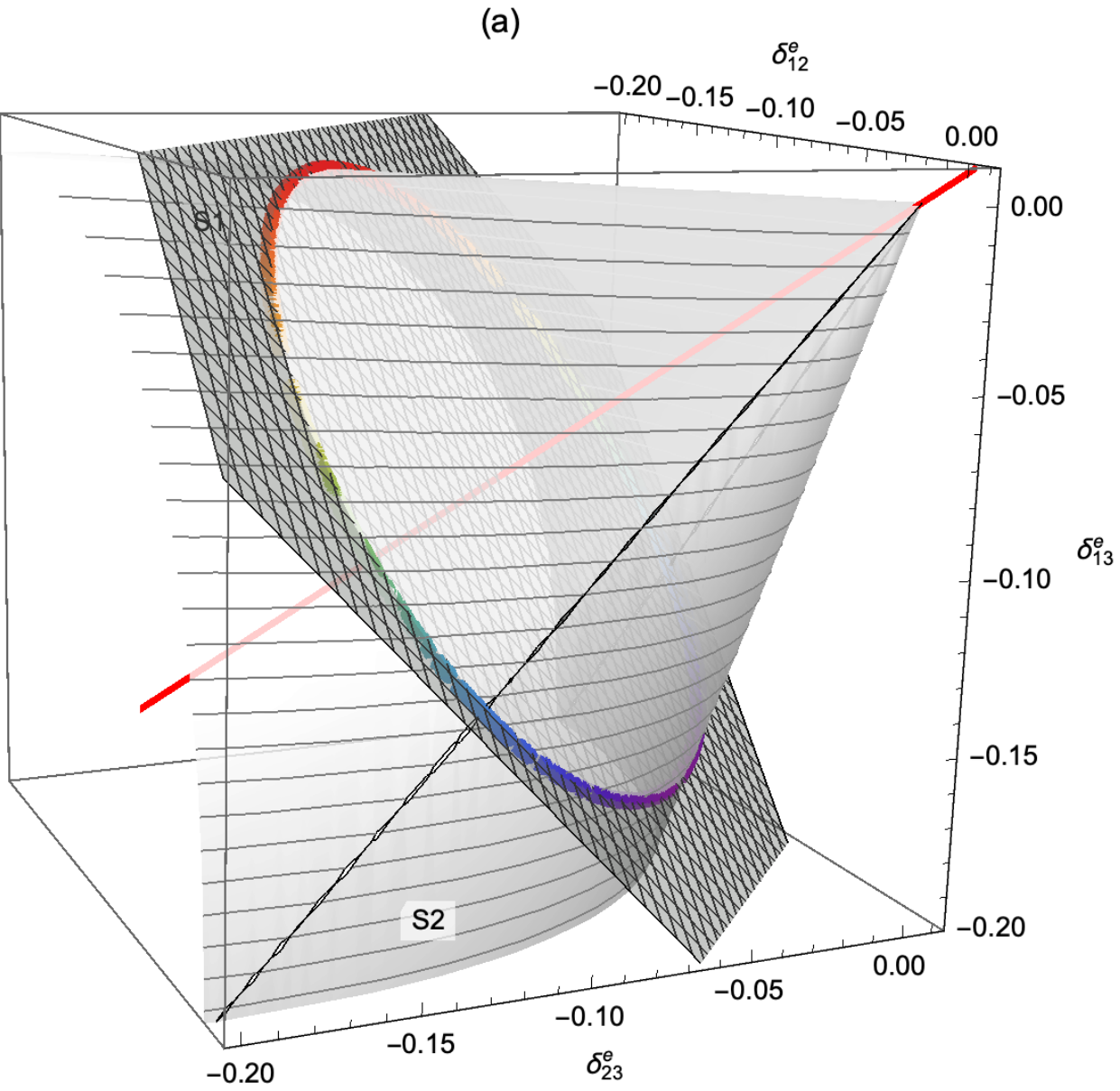}  
\includegraphics[height=0.33 \textheight]{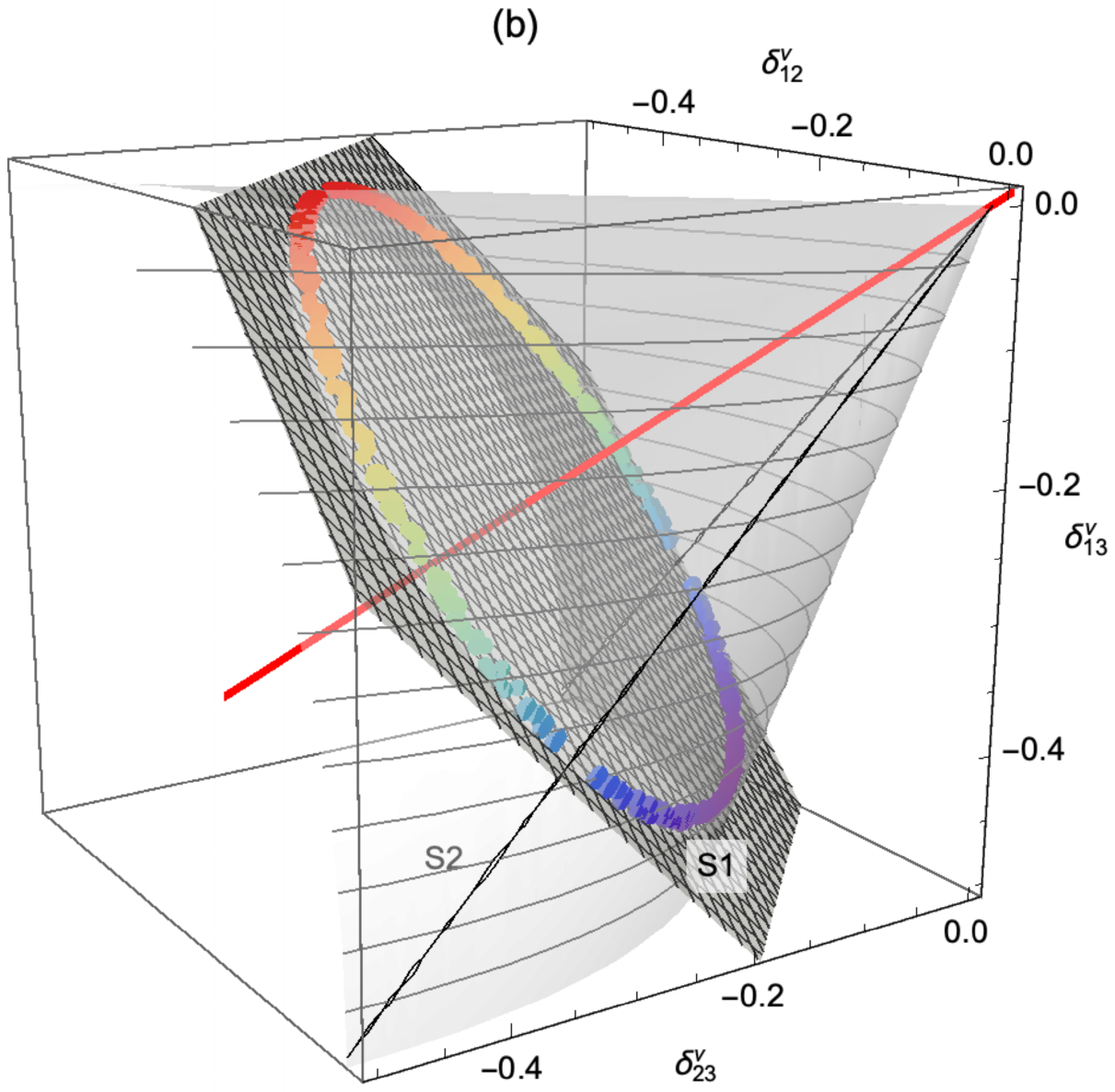}  
\caption{Parameter space for (a) charged lepton and (b) neutrinos}
\label{fig.memnu}
\end{center}
\end{figure}

Although absolute neutrino masses remain unknown, they can be calculated with the help of the mass-squared difference $\Delta m^2_{ij}$ after initializing the lightest $m^\nu_1$. 
Another unknown problem is neutrino mass order. In the close-to-flat mass structure, flavor symmetry is broken by perturbations $\delta_{ij}^f$, which is in favor of the NH neutrino. For the inverted hierarchy neutrino, nondiagonal corrections cannot be treated as perturbations.
We treat neutrino mass as a precise value and neglect the error. Charged lepton mass data are also measured with high precision. This means that there is a very thin distribution near surface $S1$ in Fig. (\ref{fig.explainVP}). Thus, we calculate lepton mass perturbations $\delta_{ij}^{\nu,e}$ from the second method. The allowed ranges are shown in Fig. (\ref{fig.memnu}). 

Until now, all quark and lepton masses have been reproduced with successful physical results. The close-to-flat mass matrix has been proven as a good mechanism to address the problematic fermion mass hierarchy. It provides a general flavor structure independent of concrete mass data. In each family, ${\bf M}_0^f$ are parameterized by 3 $\delta_{ij}^f$. The two quantities $S^f$ and $Q^f$ are generated as $h_{12}^f$ and $h_{23}^f$. The role of the remaining one (chosen as $\delta_{13}^f$) will be discussed in flavor mixing.

\section{Yukawa phase as a quantum origin of CP violation}
In addition to fermion mass, the flavor structure must address fermion mixings. 
Experiments have shown that a nonvanishing CPV exists in both CKM and PMNS mixings. The necessary condition of CPV is the existence of a complex phase in weak charge current interactions. In terms of eq. (\ref{eq.CKM}), it must come from the left-handed unitary matrix. In this section, we discuss ${\bf F}_L^f$, including its parameterization, explanation and verification.
\subsection{The parameterization of ${\bf F}_{L,R}^f$}
 Let us consider the number of d.o.f. in flavor structure. Real transformation ${\bf U}_{0}^f$ in eq. (\ref{eq.doublesideU0}) is determined by 3 nondiagonal vacuum perturbations $\delta_{ij}^f$. Two of the $\delta_{ij}^f$ generate family mass hierarchies, and the remaining $\delta_{ij}^f$ remains free in each family. 
In the CKM, 4 free parameters are needed to correspond to 3 flavor mixing angles and 1 Dirac CPV. Thus, in addition to two remaining free parameters in ${\bf I}_\Delta^u$ and ${\bf I}_\Delta^d$, only two parameters in ${\bf F}_L^{u}({\bf F}_L^{d})^\dag$ are needed for flavor mixings.
It is noted that a rephasing process still exists in mass eigenstates. In mass eigenstates labeled by $\psi^{f,m}_{L,R}$, fermion rephasing can be taken as
\begin{eqnarray}
	{\psi}^{u,m}_{L,R}\rightarrow {\bf K}^u \psi^{u,m}_{L,R},~~~
	{\psi}^{d,m}_{L,R}\rightarrow {\bf K}^d \psi^{d,m}_{L,R},~~~
\end{eqnarray}
with diagonal
		\begin{eqnarray}
			{\bf K}^u&\equiv&{\rm diag}(e^{i\beta_1},e^{i\beta_2},e^{i\beta_3})
			\\
			{\bf K}^d&\equiv&{\rm diag}(1,e^{i\alpha_1},e^{i\alpha_2})
		\end{eqnarray}
Under the rephasing, the mass terms remain invariant, but the CKM matrix can be transformed into the standard from 
\begin{eqnarray}
	{\bf U}_{CKM}\rightarrow {\bf K}_u^\dag {\bf U}_{CKM}{\bf K}_d.
\end{eqnarray}
(See Appendix B for the details of calculation of the rephasing matrix.) Using the rephasing, ${\bf F}^{f}_{L,R}$ can be eliminated in a global phase and parameterized as follows:
\begin{eqnarray}
{\bf {F}}^d_L={\rm diag}(1,1,1),~~~
{\bf {F}}^u_L={\rm diag}(1,e^{i\lambda^u_1},e^{i\lambda^u_2}).
\end{eqnarray}
Substituting them into eq. (\ref{eq.Mdecomposition}), the quark mass matrix can be written in the form
	\begin{eqnarray}
	{\bf M}^d&=& \frac{v_0}{3\sqrt{2}}y^d{\bf I}_\Delta^d=\frac{v_0}{3\sqrt{2}}y^d\left(\begin{array}{ccc}1 & 1+\delta_{12}^d & 1+\delta_{13}^d\\ 1+\delta_{12}^d & 1 & 1+\delta_{23}^d \\ 1+\delta_{13}^d & 1+\delta_{23}^d & 1\end{array}\right),
	\nonumber\\
	{\bf M}^u&=& \frac{v_0}{3\sqrt{2}}y^u({\bf F}_L^u)^\dag {\bf I}_\Delta^u= \frac{v_0}{3\sqrt{2}}y^u\left(\begin{array}{ccc}1 & 1+\delta_{12}^u & 1+\delta_{13}^u\\ (1+\delta_{12}^u)e^{-i\lambda^u_1} & e^{-i\lambda^u_1} & (1+\delta_{23}^u)e^{-i\lambda^u_1} \\ (1+\delta_{13}^u)e^{-i\lambda^u_2}& (1+\delta_{23}^u)e^{-i\lambda^u_2} & e^{-i\lambda^u_2}\end{array}\right).
	\end{eqnarray}
Additionally, the CKM mixing has the form
	\begin{eqnarray}
	{\bf U}_{CKM}= {\bf U}_0^u \text{~diag}\left(1,e^{i\lambda^u_1},e^{i\lambda^u_2}\right)({\bf U}_0^d)^T
	\end{eqnarray}
Real orthogonal transformations ${\bf U}_0^{u,d}$ are completely determined by perturbations $\delta_{ij}^{u,d}$ from eq. (\ref{eq.quarkU0}).
Up to now, we have parameterized the fermion mass matrix into ten free parameters, for which roles in flavor structure are summarized in Tab. (\ref{tab.flavorstructure}).
\begin{table}[htp]
\begin{center}
\begin{tabular}{ccccc}
\hline\hline
structure && parameters && observables 
\\
\hline
$\begin{array}{c}
	\\
	{\bf I}_\Delta^u,~{\bf I}_\Delta^d
	\\
	\\
	{\bf F}_L^u
	\\
	{\bf F}_L^d={\bf 1}
	\\
	\end{array}$
&\hspace{-0.1cm}$\left.\begin{array}{c}
		\left\{\begin{array}{c}
		\\ \\ \\
		\end{array}\right.
		\\
		\left\{\begin{array}{c}
		\\ \\ 
		\end{array}\right.
	\end{array}\right.$
\hspace{-0.6cm}
&$\begin{array}{c}
		\begin{array}{cc}
		\delta_{12}^d,&\delta_{12}^u
		\\
		\delta_{23}^d,&\delta_{23}^u
		\\
		\delta_{13}^d,&\delta_{13}^u
		\end{array}
		\\
		\begin{array}{c}
			\lambda^u_1,\lambda^u_2
			\\
			-
		\end{array}
	\end{array}$
&\hspace{-0.4cm}$\left.\begin{array}{c}
		\left.\begin{array}{c}
		\\ \\
		\end{array}\right\}
		\\
		\left.\begin{array}{c}
		\\ \\ \\
		\end{array}\right\}
	\end{array}\right.$
\hspace{-0.4cm}
&$\left.\begin{array}{c}
		\left.\begin{array}{cc}h_{12}^d,&h_{12}^u
		\\
		h_{23}^d,&h_{23}^u
		\end{array}\right\} \text{4 mass hierarchies}
		\\
		\left.\begin{array}{cc}
		\theta_{12},
		\\
		\theta_{23},& \delta_{CP},
		\\
		\theta_{13},
		\end{array}\right\} \text{3 mixing angles and 1 CPV}
	\end{array}\right.$
\\
\hline
&&$y^u,~y^d$&& $\sum_i m_i^u,\sum_i m_i^d$ : \text{2 total family masses} 
\\
\hline
\hline
\end{tabular}
\end{center}
\caption{Parameters in the minimal quark flavor structure}
\label{tab.flavorstructure}
\end{table}%
Similarly, lepton unitary matrixes ${\bf F}_{L,R}^{\nu,e}$ are set as 
\begin{eqnarray}
{\bf {F}}^\nu_L=diag(1,e^{i\lambda_1^\nu},e^{i\lambda^\nu_2}),~~~
{\bf {F}}_L^e=diag(1,1,1).
\end{eqnarray}
The charged lepton mass structure can be expressed as
	\begin{eqnarray}
	{\bf M}^e&=& \frac{v_0}{3\sqrt{2}}y^e{\bf I}_\Delta^e=\frac{v_0}{3\sqrt{2}}y^e\left(\begin{array}{ccc}1 & 1+\delta_{12}^e & 1+\delta_{13}^e\\ 1+\delta_{12}^e & 1 & 1+\delta_{23}^e \\ 1+\delta_{13}^e & 1+\delta_{23}^e & 1\end{array}\right).
	\end{eqnarray}
	\begin{eqnarray}
	{\bf M}^\nu&=& \frac{v_0}{3\sqrt{2}}y^\nu({\bf F}_L^\nu)^\dag {\bf I}_\Delta^\nu= \frac{v_0}{3\sqrt{2}}y^\nu\left(\begin{array}{ccc}1 & 1+\delta_{12}^\nu & 1+\delta_{13}^\nu\\ (1+\delta_{12}^\nu)e^{-i\lambda_1^\nu} & e^{-i\lambda_1^\nu} & (1+\delta_{23}^\nu)e^{-i\lambda_1^\nu} \\ (1+\delta_{13}^\nu)e^{-i\lambda_2^\nu}& (1+\delta_{23}^\nu)e^{-i\lambda_2^\nu} & e^{-i\lambda^\nu_2}\end{array}\right)
	\end{eqnarray}
\begin{eqnarray}
\end{eqnarray}
The PMNS mixings are determined by real orthogonal matrix ${\bf U}_0^{e,\nu}$ as
	\begin{eqnarray}
		{\bf U}_{PMNS}
		&=&{\bf U}_0^e \text{~diag}\left(1, e^{-i\lambda^\nu_1}, e^{-i\lambda^\nu_2}\right)({{\bf U}_0^\nu})^{T}.
	\end{eqnarray}
\subsection{Yukawa phase}
Now, let us analyze the physical means of ${\bf F}_{L,R}^{f}$. In eq. (\ref{eq.CKM}), ${\bf {F}}^u_L({\bf {F}}_L^d)^\dag$ provides the origin of CPV in flavor mixing. 
Using eqs. (\ref{eq.diagonalizetransf}) and (\ref{eq.Mdecomposition}), the quark mass term is	\begin{eqnarray}
		-\mathcal{L}_M^f
		&=& \bar{d}_L({\bf F}_{L}^d)^\dag {\bf M}_0^d  {\bf F}_{R}^d d_R
			+\bar{u}_L({\bf F}_{L}^u)^\dag {\bf M}_0^u {\bf F}_{R}^u u_R+H.c.
	\end{eqnarray}
Defining Yukawa interaction eigenstates as
\begin{eqnarray}
d_{L,R}^{(Y)}={\bf F}^d_{L,R} d_{L,R},~~~
u_{L,R}^{(Y)}={\bf F}^u_{L,R} u_{L,R},~~~
Q_L^{(Y)}=(u_L^{(Y)}, d_L^{(Y)})^T
\label{eq.defineYukawaPhase}
\end{eqnarray}
the above mass term is written as
	\begin{eqnarray}
		-\mathcal{L}_M^q&=& \frac{1}{\sqrt{2}}\frac{{y}^d}{3}\bar{d}_L^{(Y)} {\bf I}_\Delta^d v_0  d_R^{(Y)}
			+\frac{1}{\sqrt{2}}\frac{{y}^u}{3}\bar{u}_L^{(Y)} {\bf I}_\Delta^dv_0 u_R^{(Y)}+H.c.
		\nonumber\\
		&=&\frac{{y}^d}{3} \bar{Q}_L^{(Y)} {\bf I}_\Delta^d\left(\begin{array}{c}0 \\ v_0/\sqrt{2}\end{array}\right)  d_R^{(Y)}
			+\frac{{y}^u}{3}\bar{Q}_L^{(Y)}{\bf I}_\Delta^u\left(\begin{array}{c} v_0/\sqrt{2}\\ 0\end{array}\right) u_R^{(Y)}+H.c.
	\label{eq.derYukawaPhase}
	\end{eqnarray}
The results can be generalized into the lepton sector. The Dirac neutrino mass and charged lepton mass can be expressed into Yukawa state $L_L^{(Y)}, \nu_R^{(Y)}, e_R^{(Y)}$ as 
	\begin{eqnarray}
		-\mathcal{L}_M^l&=& \frac{{y}^e}{3}\bar{L}_L^{(Y)} {\bf I}_\Delta^e\left(\begin{array}{c}0 \\ v_0/\sqrt{2}\end{array}\right)  e_R^{(Y)}
			+\frac{{y}^\nu}{3}\bar{L}_L^{(Y)}  {\bf I}_\Delta^\nu\left(\begin{array}{c} v_0/\sqrt{2}\\ 0\end{array}\right) \nu_R^{(Y)}+H.c.
		\label{eq.YukawaEigenStatesLepton}
\label{eq.derYukawaPhase}
\end{eqnarray}
Above results show some organized flavor structure with real couplings in the new Yukawa states. Complex Yukawa phases required by non-vanishing CPV are introduced as superposition coefficients between Yukawa states and weak gauge states.
The unitary matrix ${\bf F}_{L,R}^f$ can be explained as a unitary transformation between two kind of representations. 
It provides a new way to understand CPV in flavor mixing from a quantum effect.

\subsection{Flavor mixing angles and CPV}
\begin{figure}[htbp]
\begin{center}
	\includegraphics[height=0.23 \textheight]{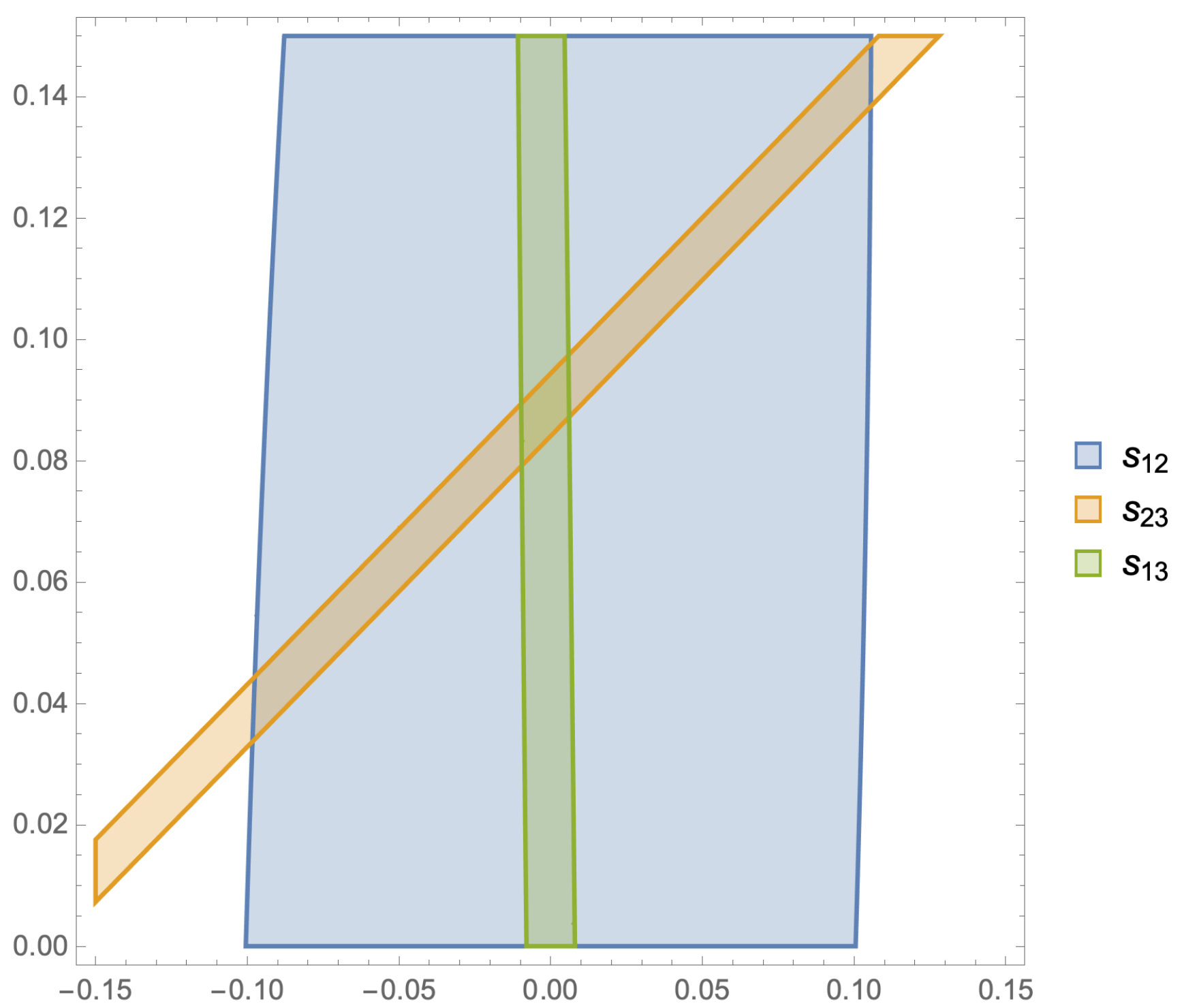}  
	\hspace{0.5cm}
	\includegraphics[height=0.23 \textheight]{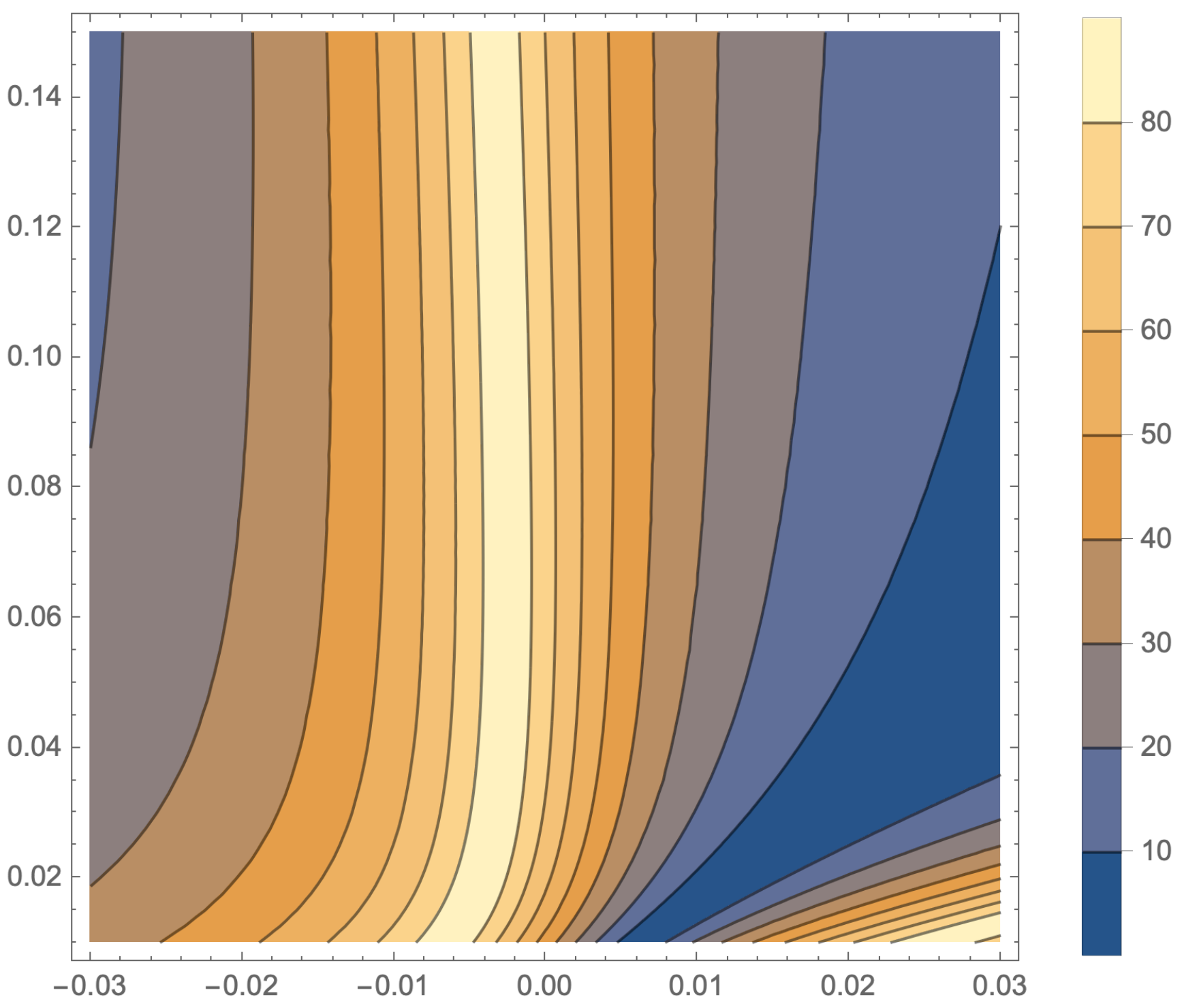}  
\caption{Constraints from three CKM mixing angles (left) at 95\% CL and contour line of CPV (right) [degree] at the matched $\delta_{ij}^{u,d}$ given in Tab. \ref{tab.quarkresult} on the $\lambda^u_1,\lambda^u_2$ plane.}
\label{fig.CKMYukawaPhase}
\end{center}
\end{figure}

\begin{figure}[htbp]
\begin{center}
	\includegraphics[height=0.23 \textheight]{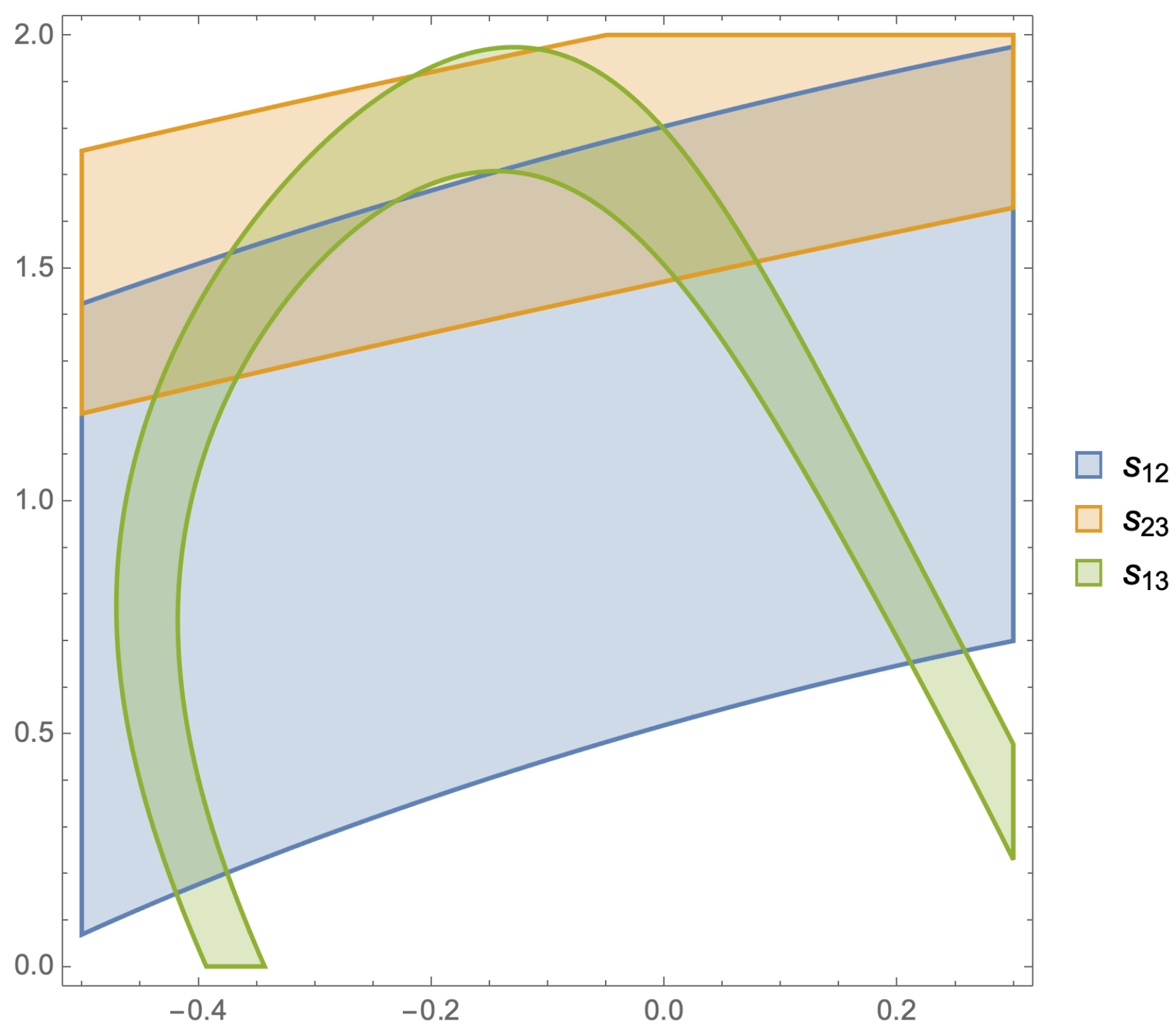}  
	\hspace{0.5cm}
	\includegraphics[height=0.23 \textheight]{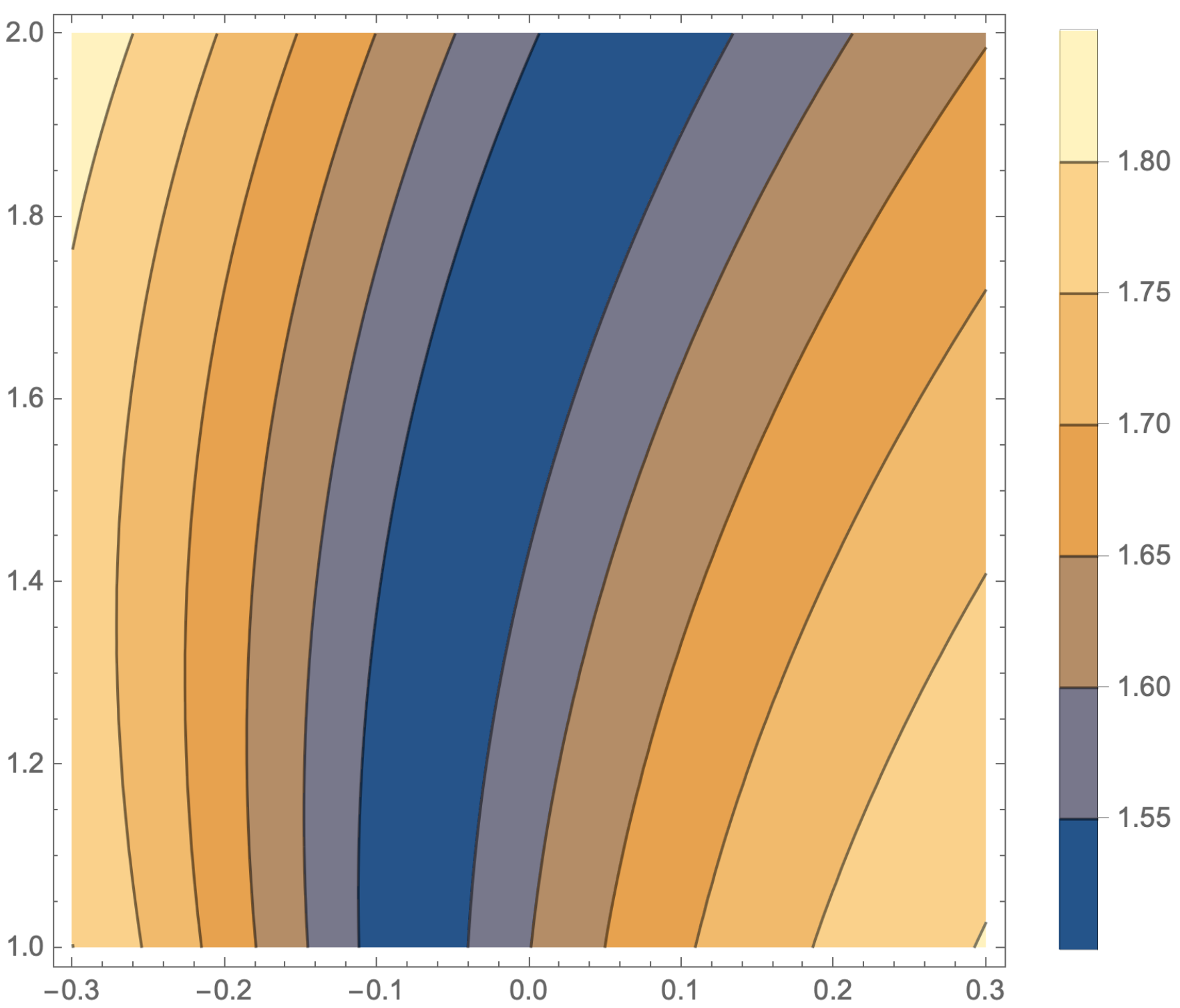}  
\caption{Constraints from three PMNS mixing angles (left) at $3\sigma$ and contour line of CPV (right) [radian/$\pi$] at the matched $\delta_{ij}^{\nu,e}$ given in Tab. \ref{tab.leptonresult} on the $\lambda^\nu_1,\lambda^\nu_2$ plane.}
\label{fig.PMNSYukawaPhase}
\end{center}
\end{figure}

In addition to parameters for mass hierarchies, the minimal flavor structure provides four remaining free parameters for flavor mixings: $\delta_{13}^{u,d}$ and Yukawa phases $\lambda_1^u,\lambda_2^u$. The next mission is to verify the validity of the minimal flavor structure with respect to quark CKM mixing.
The process is operated as follows: 
(1) choose a random point $(\delta_{12}^d,\delta_{23}^d,\delta_{13}^d)$ from Fig. 2(a) and another $(\delta_{12}^u,\delta_{23}^u,\delta_{13}^u)$ from Fig. 2(b);
(2) calculate $U_{CKM}$ by scanning all possible Yukawa phases $\lambda^u_1$ and $\lambda_2^u$ in terms of eqs. (\ref{eq.calS123}, \ref{eq.calCPV}) in Appendix B;
(3) if the mixing results agree with the CKM data, record $(\delta_{12}^d,\delta_{23}^d,\delta_{13}^d)$ and   $(\delta_{12}^u,\delta_{23}^u,\delta_{13}^u)$ as a set of acceptable results. 
After tedious numerical computation, a matched result is listed in Tab. (\ref{tab.quarkresult}). (Alternatively, Wolfenstein parameterization can be defined in \cite{WolfensteinPRL1983}.)
Up until now, all current data, quark masses and CKM mixings have been reproduced successfully.  
At the matched points, allowed Yukawa phase space is shown in Fig. \ref{fig.CKMYukawaPhase}.
It is also effectively checked by operating the same matching process as for PMNS. 
Setting $m^\nu_1=0.0001~eV$ as an example, the matched results are listed in Tab. \ref{tab.leptonresult}, and the allowed Yukawa phase is shown in Fig. \ref{fig.PMNSYukawaPhase}.
\begin{table}[htp]
\begin{center}
\begin{tabular}{c|c|c|c}
\hline\hline
&$\delta^d_{ij}$ & $\delta^u_{ij}$ & $ \lambda^u_i$ 
\\
\hline
para. &$\begin{array}{c}\delta_{12}^d=-0.00723\\
	\delta_{23}^d=-0.0644\\
	\delta_{13}^d=-0.0377\end{array}$ 
   & $\begin{array}{c}\delta_{12}^u=-0.0000453\\
	\delta_{23}^u=-0.0172\\
	\delta_{13}^u=-0.0165\end{array}$
  & $\begin{array}{c}
	\lambda_1^u=-0.00504\\
	\lambda_2^u=0.0851
	\end{array}$
\\
\hline
\hline
&$m^d_i$ & $m^u_i$ & $ CKM$ 
\\
\hline
results &   $\begin{array}{c}m^d=4.750~\textrm{MeV}\\
	m^s=98.96~\textrm{MeV}\\
	m^b=4.176~\textrm{GeV}
	\end{array}$
   & $\begin{array}{c}m^u=2.205~\textrm{MeV}\\
	m^c=1.303~\textrm{GeV}\\
	m^t=173.0~\textrm{GeV}
	\end{array}$
  & $\begin{array}{c}s_{12}=0.2243\\
	s_{23}=0.04141\\
	s_{13}=0.003942\\
	\delta_{CP}=75.07^\circ\end{array}$
\\
\hline\hline
\end{tabular}
\end{center}
\caption{The matched results from quark flavor structure. $v_0y^f/\sqrt{2}$ is set as the total family mass of quarks.}
\label{tab.quarkresult}
\end{table}%
\begin{table}[htp]
\begin{center}
\begin{tabular}{c|c|c|c}
\hline\hline
&$\delta^e_{ij}$ & $\delta^\nu_{ij}$ & $ \lambda^\nu_i$ 
\\
\hline
para. &$\begin{array}{c}
	\delta_{12}^e=-0.00443\\
	\delta_{23}^e=-0.146\\
	\delta_{13}^e=-0.105\end{array}$ 
   & $\begin{array}{c}
   	\delta_{12}^\nu=-0.176\\
	\delta_{23}^\nu=-0.435\\
	\delta_{13}^\nu=-0.073\end{array}$
  & $\begin{array}{c}
	\lambda_1^\nu=-0.0111\\
	\lambda_2^\nu=1.60\end{array}$
\\
\hline
\hline
&$m^e_i$ & $m^\nu_i$ & $ PMNS$ 
\\
\hline
results &   $\begin{array}{c}
	m^e=0.5108~\textrm{MeV}\\
	m^\mu=105.55~\textrm{MeV}\\
	m^\tau=1.777~\textrm{GeV}
	\end{array}$
   & $\begin{array}{c}
   	m^\nu_1=0.0001~\textrm{eV}\\
	m^\nu_2=0.008640~\textrm{eV}\\
	m^\nu_3=0.04996~\textrm{eV}
	\end{array}$
  & $\begin{array}{c}s^2_{12}=0.3353\\
	s^2_{23}=0.4393\\
	s^2_{13}=0.02009\\
	\delta_{CP}=1.486\pi\end{array}$
\\
\hline\hline
\end{tabular}
\end{center}
\caption{The matched results from leptonic flavor structure.$v_0y^f/\sqrt{2}$ is set as the total family mass of leptons}
\label{tab.leptonresult}
\end{table}

The numerical results in Tab. \ref{tab.quarkresult} and \ref{tab.leptonresult} show $\lambda^f_1\ll \lambda^f_2$ for $f=u,\nu$. It hints that an approximation relation between weak gauge states and Yukawa interaction states
		\begin{eqnarray*}
		&& u^{(Y)}_1=u_1,~~u^{(Y)}_2\simeq u_2,~~ u^{(Y)}_3=e^{-i\lambda^u_2}u_3,
		\\
		&& \nu^{(Y)}_1=\nu_1,~~\nu^{(Y)}_2\simeq\nu_2,~~\nu^{(Y)}_3=e^{i\lambda^\nu_2}\nu_3.
		\end{eqnarray*}
In the case, the flavor mixing is dominated by three  parameters: $\delta_{13}^u,\delta_{13}^d,\lambda_2^u$ for CKM ($\delta_{13}^\nu,\delta_{13}^e,\lambda_2^\nu$ for PMNS). If an underlying physics mechanism exists behind the approximation relation, a strong connection between mixing angles and CPV will be implied. This issue would be the important direction for future researches.
\section{Flavor Models}
The minimal flavor structure must be realized by parameterization of real ${\bf M}_0^f$ and left-handed unitary phase ${\bf F}_{L,R}^f$ in each family.  
All fermion masses, flavor mixing angles and CPV have been reproduced successfully. 
In this section, we discuss some possible phenomenological scenarios that can yield the minimal flavor structure. 
In terms of the origin of close-to-flat structure, the flavor models have primarily been of three types:
\begin{itemize}
	\item[1.] Quasi-universal interaction model. In this case, close-to-flat flavor structure comes from Yukawa couplings. This pattern indicates quasi-family-universal fermion couplings to Higgs. Only slight deviations appear at nondiagonal element $\delta_{ij}^f$. The model is still based on the SM Higgs doublet, in which FCNC does not appear in the tree level. The quark Yukawa terms can be expressed by
	\begin{eqnarray}
		-\mathcal{L}^q_Y= \frac{y^d}{3}({\bf I}_\Delta^d)_{ij}\bar{Q}_L^{(Y),i} Hd_R^{(Y),j}
		+\frac{y^u}{3}({\bf I}_\Delta^u)_{ij}\bar{Q}_L^{(Y),i} \tilde{H} u_R^{(Y),j}+H.c.
	\end{eqnarray}
The quasi-democratic model in \cite{Sogami1998PTP,Miura2000} proposes a similar mass matrix, in which the pattern was expressed by a unitarity matrix with complex nondiagonal perturbations. The difference is that we  need only three real symmetric perturbations to replace complex nondiagonal corrections in the democratic-type model. The origin of CPV in our paper comes from quantum phases.
	\item[2.] Multi-Higgs models. In this case, ${\bf I}_\Delta^f$ comes not from couplings but from the vacuum structure. 
Nondiagonal flavor breakings can be fulfilled in the context of an extension of the SM, where some Higgs doublets and/or even more multiplets are introduced \cite{Botella2010PLB,Haba2010EPJC}. 
Generally, more than two Higgs scalars are needed for CPV in flavor mixing.
By appropriate assignment of VEVs of scalars and Yukawa couplings, it is possible to obtain close-to-flat vacuum. More generally, in multiscalar-inspired scenarios, couplings between scalars and fermions can result in special flavor structures after scalar gain of corresponding VEVs, which is the case in so-called flavon models \cite{Bazzocchi2004PRD,BazzocchiPRD2004,PascoliJHEP2016}.
For example, let us consider a flavon model with three flavons $\phi_{i}$ ($i=1,2,3$) and an assisted flavon $\phi_0$ as SM gauge group singlets and flavor triplets.
In the flavor space, flavons obtain their VEVs as
	\begin{eqnarray*}
		\langle\phi_1\rangle=\left(\begin{array}{c}1\\ 0\\ 0\end{array}\right),~
		\langle\phi_2\rangle=\left(\begin{array}{c}0\\ 1\\ 0\end{array}\right),~
		\langle\phi_3\rangle=\left(\begin{array}{c}0\\ 0\\ 1\end{array}\right),~
		\langle\phi_0\rangle=\left(\begin{array}{c}1\\ 1\\ 1\end{array}\right).	\end{eqnarray*}
The flavor structure can be generated from SM-like terms in Yukawa basis
	\begin{eqnarray*}
	-\mathcal{L}_Y^q&=&\frac{y^u}{3}\bar{Q}_L^{(Y)}\phi_0\tilde{H}\phi_0^T u_R^{(Y)}+\frac{y^d}{3} \bar{Q}_L^{(Y)}\phi_0{H}\phi_0^Td_R^{(Y)}
		\\
		&&+\sum_{i\neq j}\frac{y^u\delta^u_{ij}}{3}\bar{Q}_L^{(Y)}\phi_i\tilde{H}\phi_j^Tu_R^{(Y)}
		+\sum_{i\neq j}\frac{y^d\delta^d_{ij}}{3} \bar{Q}_L^{(Y)}\phi_i{H}\phi_jd_R^{(Y)}+H.c.
	\end{eqnarray*} 
with perturbation $\delta^f_{ij}=\delta^f_{ji}$ ($i,j=1,2,3$ and $i\neq j$) and the SM Higgs doublet $H$.
In the scenario, the flat flavor structure is generated by $\phi_0$ and flavor breaking is achieved by $\phi_i$. Complex phases required by CP violation is still provided from a quantum superposition between gauge basis and Yukawa basis.
	\item[3.] There is another possible mechanism to provide the flavor structure. When considering flavor interaction dependence on energy scale, the small broken flavor effects may result from the flat Higgs vacuum and/or Yukawa couplings by a renormalization effect \cite{ZXingPRD2001}. 
If the mechanism is confirmed in future work, the flavor puzzle will be self-determined in the SM.  
\end{itemize}
These possible scenarios provide a wide range of new physics.
Regardless of the scenario, a (quasi-)family universal Yukawa coupling is always hinted as a favorable result.
\section{Conclusion}
The SM fermion flavor struture has been analyzed by bi-unitary decomposition of mass matrix. The real matrix ${\bf M}_0^f$ and unitary ${\bf F}_{L}^f$ play different roles. The former generates fermion mass hierarchies, and the latter provides the CPV in flavor mixing. The minimal flavor structure has been realized by close-to-flat real symmetric ${\bf I}_\Delta^f$. The number of parameters in the minimal flavor structure is equal to phenomenological observables, six fermion masses and four mixings, without any redundancy. The structure successfully reproduces current mass and mixing data in both quark sector and lepton sector. An important indication from the minimal flavor structure is that the origin of CPV existing in CKM and PMNS may be attributed to a quantum effect. The complex phases can arise from the superposition between Yukawa interaction states and weak gauge states. This information helps us understand the properties of Yukawa interactions in a new way. In future research, the possible mechanisms or models corresponding to ${\bf I}_\Delta^f$ merit further investigation, which will reveal the nature of flavor.

\section*{Acknowledgements}
I thank my collaborator Prof. Rong Li for helpful discussions on the subject. This work is partially supported  by the Fundamental Research Funds for the Central Universities.

\begin{appendix}
\section{Data of Fermion Masses and Mixings}
All data of quark and charged lepton masses as well as CKM and PMNS mixings come from \cite{2018PDG}. In Tab. IV, we set $m^\nu_1=0.0001$~eV, and other neutrino masses can be calculated from the mass squared differences $\Delta m^2_{ij}$.
\begin{table}[htp]
\begin{center}
\caption{Quark and lepton masses}
\begin{tabular}{|c|}
\hline
\hline
quark mass 
\\
\hline
$\begin{array}{ll}
	m_u=2.2^{+0.5}_{-0.4}~\textrm{MeV}
	&
    	m_d=4.7^{+0.5}_{-0.3}~\textrm{MeV}
	\\
	m_c= 1.275^{+0.025}_{-0.035}~\textrm{GeV}
	&
	m_s=95^{+9}_{-3}~\textrm{MeV}
	\\
	m_t= 173.0\pm0.4~\textrm{GeV}
	&
	m_b=4.18^{+0.04}_{-0.03}~\textrm{GeV}
	\end{array}$
	\\
\hline
 lepton mass
 \\
 \hline
	$\begin{array}{ll}
	m_e=0.5109989461\pm0.0000000031~\textrm{MeV}
	&
	m_1^\nu=0.0001~\textrm{eV} (input)
	\\
	m_\mu= 105.6583745\pm 0.0000024~\textrm{MeV}
	&
	m_2^\nu=0.0086~\textrm{eV}
	\\
	m_\tau= 1776.86\pm 0.12~\textrm{MeV}
	&
	m_3^\nu=0.050~\textrm{eV}
	\end{array}$
\\
\hline\hline
\end{tabular}
\end{center}
\label{tab.quarkleptonmassdata}
\end{table}%
\begin{table}[htp]
\begin{center}
\caption{CKM and PMNS experimental data}
\begin{tabular}{|c|c|}
\hline
\hline 
CKM & PMNS,~$3\sigma$(2018PDG)
\\
\hline
	$\begin{array}{l}
	s_{12}=\frac{|V_{us}|}{\sqrt{|V_{ud}|^2+|V_{us}|^2}}=0.2244\pm0.0005
	\\
	s_{23}=\frac{|V_{cb}|}{\sqrt{|V_{ud}|^2+|V_{us}|^2}}=0.0422\pm0.0008
	\\
	s_{13}=|V_{ub}|=0.00394\pm0.00036
	\\
	\delta=(73.5^{+4.2}_{-5.1})^\circ
	\end{array}$
&
	$\begin{array}{l}
	s_{12}^2=0.297,~~0.250-0.354
	\\
	s_{23}^2=0.425,~~0.381-0.615
	\\
	s_{13}^2=0.0215(NH),~0.0190-0.0240
	\\
	\delta=1.38\pi,~~2\sigma: (1.0-1.9)\pi
	\end{array}$
\\
\hline\hline
\end{tabular}
\end{center}
\label{tab.flavormixingdata}
\end{table}%
\section{Calculation of Flavor Mixing Angles and CPV}
After bi-unitary transformation eq. (\ref{eq.diagonalizetransf}), the diagonalized quark masses and weak charge current terms are
\begin{eqnarray}
 \mathcal{L}_W^q=\bar{u}^m_L{\bf m}^u u^m_R+\bar{d}^m_L{\bf m}^d d^m_R
 	+\frac{g}{\sqrt{2}}\bar{u}^{m}_L\gamma^\mu{\bf U}_{CKM}d^{m}_L W_\mu
			+H.c.
\end{eqnarray} 
There is still a rephasing transformation to keep the diagonal mass invariant:
\begin{eqnarray}
	{u}^m_{L,R}\rightarrow {\bf K}^u u^m_{L,R},~~~
	{d}^m_{L,R}\rightarrow {\bf K}^d d^m_{L,R}
\end{eqnarray}
with 
	\begin{eqnarray}
		{\bf K}_u&\equiv&{\rm diag}(e^{i\beta_1},e^{i\beta_2},e^{i\beta_3})
		\\
		{\bf K}_d&\equiv&{\rm diag}(1,e^{i\alpha_1},e^{i\alpha_2}).
	\end{eqnarray}
However, the CKM matrix transform is
\begin{eqnarray}
	{\bf U}_{CKM}\rightarrow {\bf K}_u^\dag {\bf U}_{CKM}{\bf K}_d.
\end{eqnarray}
By choice of rephasing phases, CKM can be expressed in the standard form
\begin{eqnarray}
	{\bf U}_{CKM}=\left(\begin{array}{ccc} c_{12}c_{13} & s_{12} c_{13} & s_{13}e^{-i\delta_{CP}} \\
		-s_{12} c_{23}-c_{12} s_{23}s_{13}e^{i\delta_{CP}} & c_{12}c_{23}-s_{12}s_{23}s_{13}e^{i\delta_{CP}} & s_{23}c_{13} \\ 
		s_{12}s_{23}-c_{12}c_{23}s_{13}e^{i\delta_{CP}} & -c_{12}s_{23}-s_{12}c_{23}s_{13}e^{i\delta_{CP}} & c_{23}c_{13}\end{array}\right).
	\nonumber
\end{eqnarray}
Alternatively, Wolfenstein parameterization can be defined as
\begin{eqnarray}
	s_{12}=\lambda,~~s_{23}=A\lambda^2,~~s_{13}e^{i\delta}=\frac{A\lambda^3(\bar{\rho}+i\bar{\eta})\sqrt{1-A^2\lambda^4}}{\sqrt{1-\lambda^2}[1-A^2\lambda^4(\bar{\rho}+i\bar{\eta})]}
\end{eqnarray}
For the sake of convenience, the non-rephased CKM matrix is labeled as ${\bf U}$.
To make ${\bf U}_{11},{\bf U}_{12}, {\bf U}_{22}, {\bf U}_{23}$ real, rephasing phases can be set as follows:
\begin{eqnarray*}
	\beta_1&=&{\rm arg}({\bf U}_{11})
	\\
	\beta_2&=&{\rm arg}({\bf U}_{23})+\alpha_2
	\\
	\beta_3&=&{\rm arg}({\bf U}_{33})+\alpha_2
	\\
	\alpha_1&=&{\rm arg}({\bf U}_{11})-{\rm arg}({\bf U}_{12})
	\end{eqnarray*} 
$\alpha_2$ and $\delta_{CP}$ can be solved from 
	\begin{eqnarray*}
	\left({\bf K}_u{\bf U}{\bf K}_d^\dag\right)_{21}={\bf U}_{CKM}\Big|_{21},
	\\
	\left({\bf K}_u{\bf U}{\bf K}_d^\dag\right)_{22}={\bf U}_{CKM}\Big|_{22}.
	\end{eqnarray*}
Ignoring these details, mixing angles can be calculated directly from 
\begin{eqnarray}
	s_{13}=|U_{13}|,~~~
	s_{23}^2=\frac{|U_{23}|^2}{1-|U_{13}|^2},~~~
	s_{12}^2=\frac{|U_{12}|^2}{1-|U_{13}|^2}
	\label{eq.calS123}
\end{eqnarray}
And the CPV can be determined by $J_{CP}$ or another rephasing invariant:
	\begin{eqnarray}
c_{12}^2c_{23}^2+s_{13}^2s_{23}^2s_{12}^2-2c_{12}s_{12}c_{23}s_{23}s_{13}\cos[\delta_{CP}]=|U_{22}|^2.
		\label{eq.calCPV}
	\end{eqnarray}
\end{appendix}

\end{document}